\documentclass[11pt]{article}

\usepackage[preprint]{acl}

\usepackage{times}
\usepackage{latexsym}

\usepackage[T1]{fontenc}

\usepackage[utf8]{inputenc}

\usepackage{microtype}
\usepackage {amsmath}
\usepackage{inconsolata}

\usepackage{graphicx}
\usepackage{adjustbox}
\usepackage{pgfplots}
\pgfplotsset{compat=1.18}

\usepackage{booktabs}

\usepackage{tabularx}
\usepackage{multirow}
\usepackage{array}
\usepackage{geometry}
\usepackage{rotating} 
\usepackage{threeparttable} 
\usepackage{adjustbox} 
\usepackage[table]{xcolor}
\usepackage{colortbl}
\usepackage{amsmath}
\definecolor{red-light}{RGB}{255,230,230}
\definecolor{red-dark}{RGB}{255,100,100}
\definecolor{orange-light}{RGB}{255,235,200}
\definecolor{orange-dark}{RGB}{255,165,0}
\definecolor{green-light}{RGB}{230,255,230}
\definecolor{yellow-dark}{RGB}{100,200,100}

\definecolor{yellow-light}{RGB}{255,255,200}  
\definecolor{yellow-dark}{RGB}{255,200,0}

\title{ClarEval: A Benchmark for Evaluating Clarification Skills of Code Agents under Ambiguous Instructions}

\author{
    Jialin Li\textsuperscript{1},
    Yuan Wu\textsuperscript{1}\footnotemark[1],
    Yi Chang\textsuperscript{1,2,3}\thanks{Corresponding authors.} \\
    \textsuperscript{1}School of Artificial Intelligence, Jilin University\\
    \textsuperscript{2}Engineering Research Center of Knowledge-Driven Human-Machine Intelligence, MOE, China\\
    \textsuperscript{3}International Center of Future Science, Jilin University\\
    \texttt{jialin24@mails.jlu.edu.cn, yichang@jlu.edu.cn, yuanwu@jlu.edu.cn}
}
\begin{document}
\maketitle
\begin{abstract}
To integrate seamlessly into real-world software engineering, Code Agents must evolve from passive instruction followers into proactive collaborative partners. However, current evaluation paradigms predominantly reward "guessing" user intent under ideal conditions, neglecting the agent's ability to align with users through dialogue—a critical trait for collaborative intelligence.
In this work, we propose a paradigm shift in evaluation to drive this transition. We introduce ClarEval, a framework designed to assess an agent's "Collaborative Quotient" by simulating the inherent ambiguity of human communication. By systematically injecting three types of realistic ambiguity (missing goals, premises, and ambiguous terminology) into standard tasks, we force agents to step out of their "generator" role and engage in requirement elicitation. To quantify this capability, we propose a metric suite led by Average Turns to Clarify (ATC) and Key Question Coverage (KQC), which measure not just the correctness of the generated code, but the efficiency and precision of the collaboration. Our experiments on eleven state-of-the-art agents reveal a stark reality: while models like GPT-5-Coder excel at coding, they often lack the strategic communication skills required for efficient partnership. ClarEval thus serves as a crucial roadmap for bridging the gap between strong coders and capable collaborators.The code is available at https://github.com/JialinLi13/ClarEval
\end{abstract}

\begin{figure}[!t]
    \centering
    \includegraphics[width=0.8\linewidth]{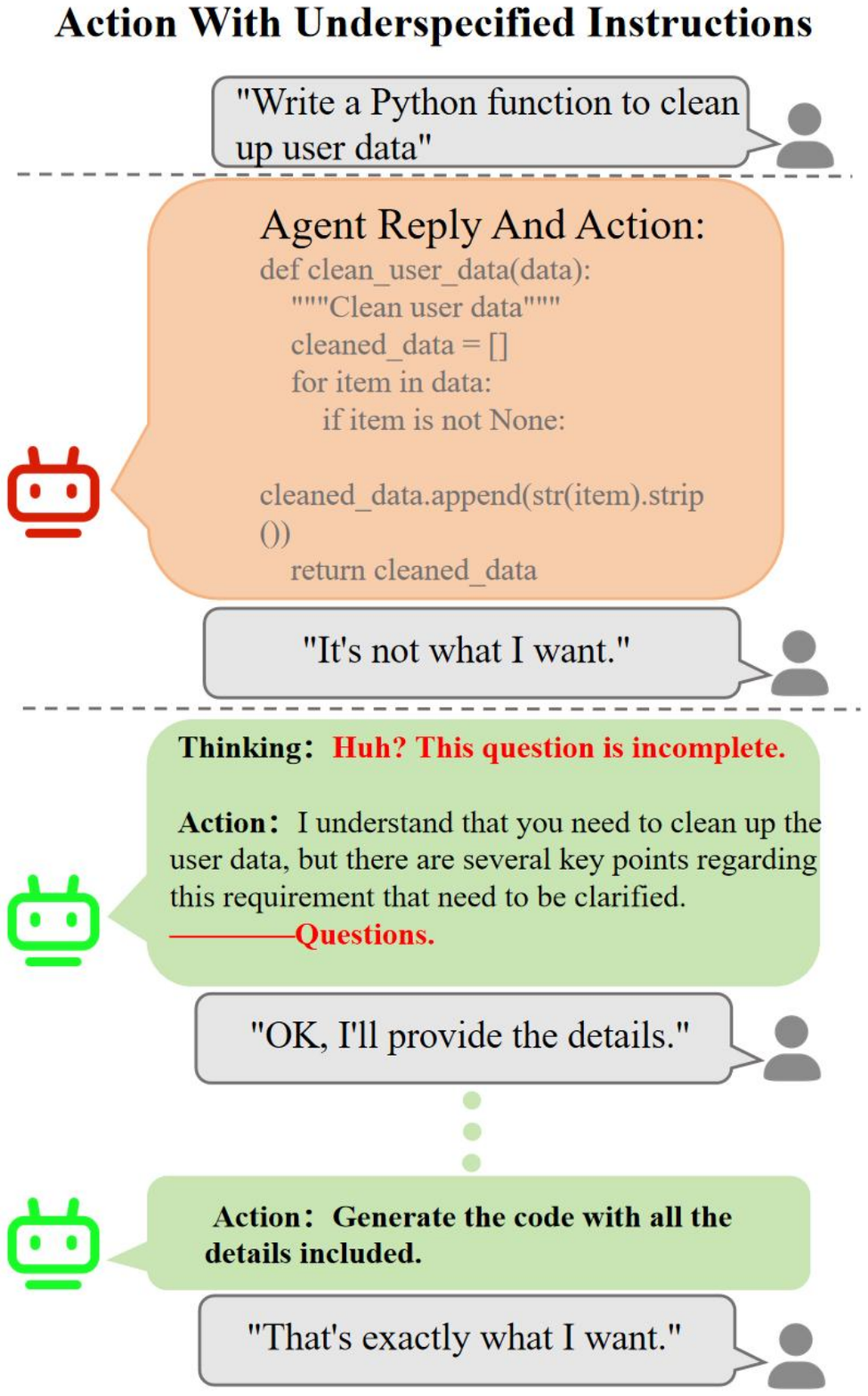}
    \caption{Contrasting agent behaviors when given an underspecified instruction. This example compares a passive generation approach, which fails due to ambiguity, with a proactive clarification approach, which successfully resolves uncertainty through dialogue before generating the correct code, highlighting the critical role of clarification skills.}
    \label{fig:enter-label}
\end{figure}

\section{Introduction}

Large Language Models (LLMs)\cite{kasneci2023chatgpt, chang2024survey, thirunavukarasu2023large} have fundamentally reshaped software engineering, powering agents capable of automating coding, debugging, and maintenance. However, the prevailing evaluation paradigm\cite{guan2025your, chen2024ppm} treats these agents as passive generators: given a static, self-contained instruction, the agent is expected to produce functionally correct code in a single turn. While this ``Pass@1'' centric approach has driven progress in syntactic correctness, it fails to capture a fundamental reality of software development: real-world requirements are rarely unambiguous or complete. In practice, instructions like ``optimize system performance'' are starting points for dialogue, not final specifications.

Current benchmarks\cite{liu2023repobench, zhuo2024bigcodebench, pan2024codev} inadvertently encourage agents to "guess" user intent to maximize pass rates, leading to misaligned outputs and eroded user trust. We argue that the next generation of Code Agents\cite{huang2023agentcoder, huang2024code} must transcend passive generation to become Active Collaborative Partners---agents capable of detecting ambiguity, inferring latent goals, and guiding users through structured clarification. However, a critical gap remains: we lack a standardized diagnostic framework to measure this ``Collaborative Quotient''. Existing evaluation paradigms either ignore interaction entirely or conflate communicative logic with the noise of complex repository contexts, making it difficult to isolate "why" an agent fails to clarify effectively.

To bridge this gap, we introduce \textbf{ClarEval}, a benchmark designed to systematically diagnose the clarification logic of Code Agents. Unlike broad-spectrum evaluations, ClarEval functions as a "unit test" for communicative intelligence. We deliberately construct our dataset by injecting controlled ambiguities---Missing Goals ($A_G$), Missing Premises ($A_P$), and Ambiguous Terminology ($A_T$)---into the canonical HumanEval dataset. While HumanEval focuses on algorithmic logic, its isolated nature allows us to assess the agent's clarification strategy in a controlled environment, free from the confounding variables of massive codebases. If an agent cannot effectively clarify a sorting order in a list, it cannot be trusted to clarify architectural decisions in a repository.

 A key contribution of ClarEval is its focus on interaction efficiency. Effective collaboration respects the user's cognitive load; an agent that asks ten vague questions is less useful than one that asks a single, precise question. We propose three targeted metrics to capture this nuance:
\begin{itemize}
    \item \textbf{Key Question Coverage (KQC):} Measures Intent Inference---the ability to identify the high-level goal hidden behind a vague prompt.
    \item \textbf{Missing Premises Recall (MPR):} Assesses Requirement Identification---ensuring the agent can surface necessary technical constraints before coding.
    \item \textbf{Average Turns to Clarify (ATC):} A novel metric quantifying Clarification Strategy. It penalizes inefficient, ``lazy,'' or redundant questioning, favoring agents that reach clarity with minimal user friction.
\end{itemize}

We evaluate eleven state-of-the-art agents using this framework. Our experiments reveal a stark reality: while models like GPT-5-Coder possess strong reasoning capabilities, most agents lack systematic clarification skills. They struggle significantly with ambiguous terminology and often resort to inefficient questioning strategies.

In summary, ClarEval establishes a standardized framework to advance Code Agents from passive code generators to proactive collaborative partners. By shifting the evaluation focus from correctness-only to collaboration-aware, we provide the necessary roadmap for developing trustworthy, human-aligned programming assistants.

\begin{table*}[htbp]
\centering
\label{tab:comparison}
\begin{adjustbox}{width=0.9\textwidth}
\begin{tabular}{lcccccc}
\toprule
\textbf{BenchMark} & \textbf{Task Type} & \textbf{Ambiguity} & \textbf{Single-round} & \textbf{Multi-round} & \textbf{Efficiency indicators} \\
\midrule
LoCoBench-Agent & software engineering & $\times$ & $\surd$ & $\surd$ & $\surd$ \\
SR-Eval & Code Generation & $\times$ & $\surd$ & $\times$ & $\times$ \\
CodeFlowBench & Code Generation & $\times$ & $\surd$ & $\surd$ & $\times$ \\
SWE-bench-Live & Issue Resolution & $\times$ & $\surd$ & $\surd$ & $\times$ \\
HumanEval & Code Generation & $\times$ & $\surd$ & $\times$ & $\times$ \\
HumanEvalComm & Code Generation & $\surd$ & $\surd$ & $\surd$ & $\times$ \\
\textbf{ClarEval} & Code Generation & $\surd$ & $\surd$ & $\surd$ & $\surd$ \\
\bottomrule
\end{tabular}
\end{adjustbox}
\caption{Benchmark comparison}
\end{table*}

\section{Related Work}

\subsection{Passive Code Generation vs. Real-World Ambiguity}
Traditional benchmarks like HumanEval \cite{chen2021evaluating}, MBPP \cite{austin2021program}, and their variants \cite{liu2023your,zhang2024humaneval} evaluate functional correctness under the assumption of ideal, self-contained specifications. This ``Pass@1'' paradigm fundamentally diverges from real-world engineering, where requirements are rarely unambiguous. In these static settings, models are incentivized to ``guess'' latent intent to maximize pass rates. Current benchmarks\cite{chen2025dynamic, hu2025dynacode，li2025refining，li2025don} lack mechanisms to penalize this guessing behavior or evaluate the agent's initiative to resolve underspecification through dialogue.

\subsection{Agentic Frameworks and Repository Context}
The field has evolved from isolated generation to autonomous agents \cite{schick2023toolformer,yang2024swe,zhang2024codeagent} operating within complex environments. Concurrently, benchmarks have scaled to the repository level \cite{wu2024repomastereval,puvvadi2025coding,chen2025acebench}, assessing code generation within broader project contexts. However, these evaluations remain predominantly execution-centric. They measure whether an agent can solve a task given a fixed context, but overlook the communicative intelligence required to elicit and define the task itself, failing to capture the collaborative dynamics essential for human-AI pairing.

\subsection{Benchmarking Clarification and Efficiency}
The evaluation of collaborative skills in code agents is an emerging frontier. Prior benchmarks for agentic software engineering largely focus on complex, multi-step implementation and real-world task success: SWE-bench-Live \cite{zhang2025swe} assesses bug-fixing robustness, while SR-Eval \cite{zhan2025sr}, CodeFlowBench \cite{wang2025codeflowbench}, and LoCoBench-Agent \cite{qiu2025locobench} evaluate multi-turn, long-context code generation under iterative or long-context requirements. However, these works do not quantify the quality or efficiency of the communication itself.

Most notably, HumanEvalComm \cite{wu2025humanevalcomm, mcnall2024humanevalcomm} pioneered the injection of ambiguity to test clarification competence. While this work successfully highlights the importance of dialogue, it focuses primarily on the binary capability of asking questions (e.g., Good-Question metric) and the ultimate correctness of the code.

ClarEval distinguishes itself by shifting the focus from "Task Success" to "Communication Efficiency and Diagnosis." (See Table 1 for a detailed comparison). We argue that an agent asking excessive or fragmented questions imposes a high cognitive tax on the user. Unlike prior works that may reward inefficient questioning, ClarEval introduces metrics like Average Turns to Clarify (ATC) and the aggregate Efficiency-Adjusted Recall (EAR) to explicitly penalize user friction. By employing a controlled ambiguity taxonomy and a deterministic simulator, ClarEval provides a precise, reproducible unit test for collaborative intelligence, prioritizing agents that achieve clarity with minimal interaction cost.

\begin{figure*}
    \centering
    \includegraphics[width=0.9\linewidth]{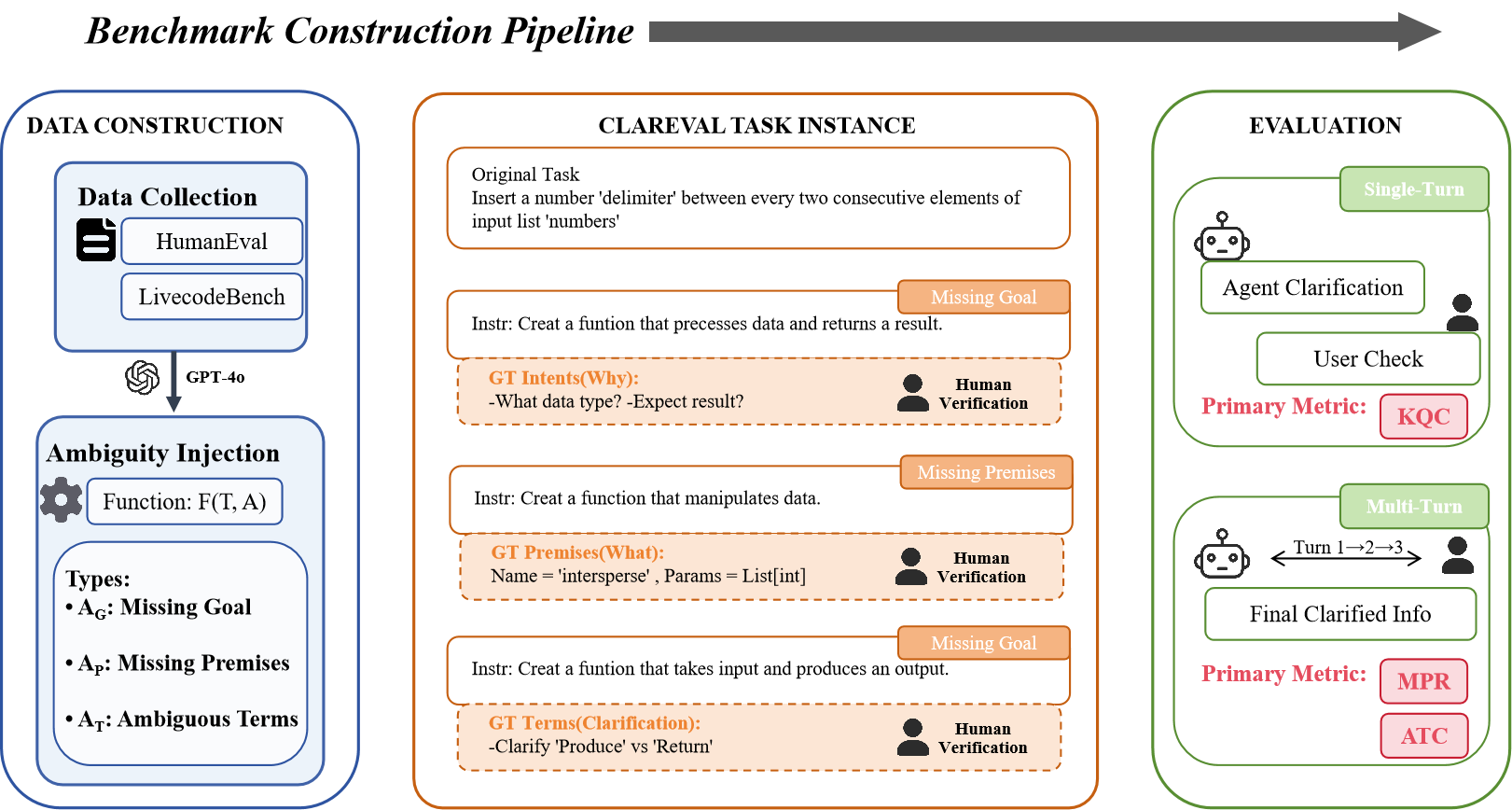}
    \caption{An overview for the ClarEval construction. First, original problems are obtained from the HumanEval dataset. Then, based on the three defined types of ambiguity—Missing Goal, Missing Premises, and Ambiguous Terminology—the original problems are transformed into ambiguous task descriptions using GPT-4o. Finally, the benchmark is refined and validated through human verification and annotation.}
    \label{fig:2}
\end{figure*}

\section{Method}

In this section, we elaborate on the construction method, evaluation scenarios, and evaluation metrics of the ClarEval benchmark. Our goal is to provide a systematic and reproducible evaluation framework for assessing the proactive clarification capability of code agents. Figure ~\ref{fig:2} overviews the construction of the ClarEval.

\subsection{Benchmark Construction}

\textbf{Source Data:} To construct a robust and contamination-aware benchmark, we integrate the HumanEval dataset~\cite{chen2021evaluating} with a challenging subset from the LiveCodeBench dataset~\cite{jain2024livecodebench}. 
The resulting collection comprises $N=750$ curated programming problems, consisting of 150 tasks from HumanEval and 600 tasks from LiveCodeBench.
We formalize this as a collection of clear source tasks $\mathcal{T} = \{T_1, T_2, \ldots, T_{N}\}$, where each task $T_i = (D_i, S_i)$ consists of a clear natural language description $D_i$ and its corresponding canonical solution $S_i$. 
By applying three distinct ambiguity injection strategies to each of the 750 source tasks, we constructed a comprehensive benchmark containing a total of 2,250 unique evaluation instances.
\textbf{Ambiguity Injection:} The core of our approach is the systematic transformation of a clear task $T_i$ into an ambiguous task $\widetilde{T}_i$. This is achieved through an ambiguity injection function $\mathcal{F}$, formally defined as:
\begin{equation}
\label{eq:injection_function}
\widetilde{T}_i = \mathcal{F}(T_i, A) = (\widetilde{D}_i, S_i)
\end{equation}
where $A \in \{A_G, A_P, A_T\}$ denotes the specific type of ambiguity applied. We define three fundamental ambiguity types:

\begin{itemize}
    \item \textbf{Missing Goal ($A_G$):} The high-level objective is omitted. Formally, if $D_i$ can be decomposed into context $C$ and goal $G$, i.e., $D_i = C \oplus G$ (where $\oplus$ denotes string concatenation), then the ambiguous description is defined as $\widetilde{D}_i = C$.
    
    \item \textbf{Missing Premises ($A_P$):} Critical technical constraints are removed. Let $\mathcal{P} = \{p_1, p_2, \ldots, p_k\}$ be the set of missing premises. The ambiguous description is generated by $\widetilde{D}_i = \text{REMOVE}(D_i, \mathcal{P})$, where $\text{REMOVE}$ eliminates all textual references to premises in $\mathcal{P}$.
    
    \item \textbf{Ambiguous Terminology ($A_T$):} A precise term $t$ is replaced with a vague term $\tilde{t}$. The transformation is defined as $\widetilde{D}_i = \text{REPLACE}(D_i, t, \tilde{t})$.
\end{itemize}

For each original task $T_i$, we construct one or more ambiguous variants $\widetilde{T}_i$ by applying $\mathcal{F}$, creating our benchmark suite. Each ambiguous task is annotated with a golden standard that defines the information required for complete disambiguation.

\subsection{Ambiguity Injection via Prompt Engineering}
The core of ClarEval lies in the systematic transformation of clear tasks into ambiguous variants. To ensure reproducibility and diversity, we employed GPT-4o as a transformation engine, guided by specific Ambiguity Injection Prompts. Instead of random token deletion, we designed three distinct prompt strategies to simulate realistic patterns of human underspecification. Table~\ref{tab:prompt_strategies} provides a comprehensive overview of these strategies and their corresponding transformation examples.

\subsection{Evaluation Settings}

We design two evaluation settings to model distinct interaction paradigms between users and code agents.

\subsubsection{Single-Turn Setting}
This setting evaluates an agent's ability to identify ambiguity and pose critical clarification questions immediately upon receiving an initial ambiguous instruction. The agent is provided only with the ambiguous task description $\widetilde{D}_i$ and generates a single response. Performance is evaluated based on the clarification questions contained in this initial reply. This setting primarily assesses the agent's initial intent inference capability (Key Question Coverage, KQC) and its zero-shot ability to infer missing premises (Single-Turn MPR), serving as a measure of initial planning quality.

\subsubsection{Multi-Turn Setting}
This setting assesses an agent's capacity for sustained dialogue, including context management, iterative guidance, and probing follow-up questions. We pre-define structured conversation scripts for each task. Starting from the ambiguous instruction $\widetilde{D}_i$, each script specifies potential user responses (which may provide partial information or introduce new ambiguities), enabling evaluation of the agent's strategic questioning throughout extended interactions.

Crucially, the evaluation of the multi-turn setting is performed using a deterministic, rule-based User Simulator script ($\mathcal{S}_i$) derived from human-annotated ground truth intents (see Appendix ~\ref{app:c} and ~\ref{sec:appendix_e}). This ensures the reproducibility and consistency of agent evaluations across different models, isolating communicative logic from dynamic simulation noise. This setting focuses on the agent's complete requirement elicitation (Dialogue-Informed Missing Premises Recall, MPR) and efficiency (Average Turns to Clarify, ATC) over the conversation . 

\subsection{Human Verification Protocol}
\label{sec:human_verification}

To ensure validity, we implemented a stringent two-stage review process involving three senior software engineers.

\paragraph{Criteria Check.} Reviewers assessed each entry based on three criteria: (1) \textbf{Ambiguity Validity}: Is the instruction genuinely unsolvable without assumptions? (2) \textbf{Plausibility}: Does the request resemble natural user prompts? (3) \textbf{Script Alignment}: Does the dialogue script accurately reflect the ground truth?

\paragraph{Ecological Validity Analysis.}
To validate the realism of synthetic ambiguity, we compared 50 ClarEval instances against real StackOverflow queries. Two experts blindly rated ``Naturalness'' on a Likert scale (1-5). Results show ClarEval's score ($4.12 \pm 0.6$) is statistically indistinguishable from real-world data ($4.25 \pm 0.8$, $p > 0.05$). This confirms that our ambiguity injection pipeline successfully mimics the ``colloquial vagueness'' found in human inquiries, ensuring the benchmark tests realistic communicative intelligence rather than synthetic pattern matching.

\paragraph{Consensus and Refinement.} We employed a majority-vote mechanism for retaining samples, achieving a Fleiss' Kappa of $\kappa = 0.82$ (substantial agreement). Ultimately, approximately 12\% of the raw generated samples were discarded or rewritten to correct hallucinations, ensuring ClarEval serves as a reliable standard.

\subsection{Evaluation Metrics}
\label{subsec:metrics}

We propose a suite of metrics to quantify clarification capability across two distinct phases: initial intent inference (single-turn) and sustained dialogue (multi-turn).

\subsubsection{Single-Turn Metrics}

These metrics evaluate the agent's immediate response to an ambiguous instruction, assessing its ability to identify the core problem without the benefit of dialogue.

\textbf{Key Question Coverage (KQC)} measures the agent's success in inferring the high-level user intent obscured by ambiguity. Let $\mathcal{K} = \{k_1, k_2, \ldots, k_m\}$ be the set of key intents required for disambiguation, as determined by expert annotation. Let $\mathcal{C} \subseteq \mathcal{K}$ be the set of intents covered by the agent's clarification questions. KQC is defined as:
\begin{equation}
\text{KQC} = \frac{|\mathcal{C}|}{|\mathcal{K}|}
\end{equation}
A higher KQC indicates superior initial intent inference capability, while discouraging agents from generating excessive, low-quality questions.

\textbf{Premise Identification Rate (PIR)} assesses the agent's zero-shot ability to identify missing technical premises from the initial ambiguous instruction. Let $\mathcal{P} = \{p_1, p_2, \ldots, p_n\}$ be the golden set of missing premises. Let $\mathcal{I} \subseteq \mathcal{P}$ be the set of premises detected in the agent's first-turn questions. PIR is defined as:
\begin{equation}
\text{PIR} = \frac{|\mathcal{I}|}{|\mathcal{P}|}
\end{equation}
PIR measures planning quality, indicating whether the agent recognizes what technical details are missing, even if it cannot yet resolve them through dialogue.

\subsubsection{Multi-Turn Metrics}

These metrics evaluate the agent's performance throughout an extended dialogue, assessing its capacity for context management, iterative questioning, and efficient collaboration.

\textbf{Key Question Coverage (KQC)} in the multi-turn setting measures the agent's success in inferring high-level user intent over the entire dialogue. Similar to the single-turn setting, let $\mathcal{K}$ be the set of key intents and $\mathcal{C}_{\text{dialogue}} \subseteq \mathcal{K}$ be the set of intents covered by the agent's questions across all turns. Multi-turn KQC is defined as:
\begin{equation}
\text{KQC} = \frac{|\mathcal{C}_{\text{dialogue}}|}{|\mathcal{K}|}
\end{equation}

\textbf{Missing Premises Recall (MPR)} evaluates an agent's completeness in resolving missing technical details over the entire dialogue. Let $\mathcal{P} = \{p_1, p_2, \ldots, p_n\}$ be the golden set of missing premises. Let $\mathcal{R} \subseteq \mathcal{P}$ be the set of premises successfully clarified by the agent by the dialogue's end. MPR is defined as:
\begin{equation}
\text{MPR} = \frac{|\mathcal{R}|}{|\mathcal{P}|}
\end{equation}
While PIR focuses on "what is missing" in a zero-shot setting, MPR focuses on "what was successfully recovered" through sustained dialogue.

\begin{table*}[htbp]
\centering

\begin{tabular}{lcccccc}
\toprule
\multirow{2}{*}{\textbf{Model}} & \multicolumn{2}{c}{\textbf{Single-Turn}} & \multicolumn{3}{c}{\textbf{Multi-Turn}} \\
\cmidrule(lr){2-3} \cmidrule(lr){4-6}
 & \textbf{KQC} & \textbf{PIR} & \textbf{KQC} & \textbf{MPR} & \textbf{ATC} \\
\midrule
Aider-GPT5        & \cellcolor{red-dark!83}0.833 & \cellcolor{orange-dark!59}0.588 & \cellcolor{red-dark!38}0.376 & \cellcolor{orange-dark!40}0.396 & \cellcolor{yellow-dark!92}1.576 \\
CodeX-GPT5        & \cellcolor{red-dark!84}0.835 & \cellcolor{orange-dark!56}0.557 & \cellcolor{red-dark!63}0.626 & \cellcolor{orange-dark!75}0.748 & \cellcolor{yellow-dark!81}1.681 \\
GPT5-Coder        & \cellcolor{red-dark!94}\underline{0.867} & \cellcolor{orange-dark!82}\underline{0.661} & \cellcolor{red-dark!53}0.529 & \cellcolor{orange-dark!95}\underline{0.951} & \cellcolor{yellow-dark!100}\underline{1.493} \\
CodeLlama-70B     & \cellcolor{red-dark!82}0.817 & \cellcolor{orange-dark!54}0.535 & \cellcolor{red-dark!43}0.427 & \cellcolor{orange-dark!39}0.385 & \cellcolor{yellow-dark!89}1.602 \\
Claude-opus4.1    & \cellcolor{red-dark!85}0.847 & \cellcolor{orange-dark!63}0.629 & \cellcolor{red-dark!75}\underline{0.754} & \cellcolor{orange-dark!93}0.930 & \cellcolor{yellow-dark!59}1.900 \\
Claude-sonnet-4.5 & \cellcolor{red-dark!85}0.852 & \cellcolor{orange-dark!63}0.631 & \cellcolor{red-dark!63}0.633 & \cellcolor{orange-dark!82}0.823 & \cellcolor{yellow-dark!72}1.719 \\
O3                & \cellcolor{red-dark!79}0.794 & \cellcolor{orange-dark!57}0.572 & \cellcolor{red-dark!57}0.573 & \cellcolor{orange-dark!79}0.791 & \cellcolor{yellow-dark!68}1.771 \\
O4                & \cellcolor{red-dark!77}0.774 & \cellcolor{orange-dark!54}0.544 & \cellcolor{red-dark!54}0.538 & \cellcolor{orange-dark!71}0.712 & \cellcolor{yellow-dark!92}1.576 \\
DeepSeek-Coder    & \cellcolor{red-dark!79}0.792 & \cellcolor{orange-dark!59}0.588 & \cellcolor{red-dark!68}0.675 & \cellcolor{orange-dark!80}0.796 & \cellcolor{yellow-dark!68}1.772 \\
Qwen2.5-Coder     & \cellcolor{red-dark!86}0.855 & \cellcolor{orange-dark!62}0.622 & \cellcolor{red-dark!54}0.544 & \cellcolor{orange-dark!64}0.643 & \cellcolor{yellow-dark!0}2.396 \\
Qwen3-Coder       & \cellcolor{red-dark!82}0.816 & \cellcolor{orange-dark!59}0.592 & \cellcolor{red-dark!67}0.668 & \cellcolor{orange-dark!64}0.640 & \cellcolor{yellow-dark!85}1.645 \\
\bottomrule

\end{tabular}
\caption{Performance comparison of code agents on ClarEval across Single-Turn and Multi-Turn settings. Metrics include KQC and MPR, where higher values indicate better clarification effectiveness, along with ATC, where lower values indicate superior efficiency. \underline{Underline} indicates the best-performing model for each metric.}
\label{tab:2}
\end{table*}

\textbf{Average Turns to Clarify (ATC)} quantifies interaction efficiency by measuring how quickly an agent secures necessary clarifications. For each clarified premise $p_i \in \mathcal{R}$, let $\text{Turn}(p_i)$ denote the dialogue turn index in which it was first successfully resolved. The ATC is calculated as:
\begin{equation}
\text{ATC} = \frac{1}{|\mathcal{R}|} \sum_{p_i \in \mathcal{R}} \text{Turn}(p_i)
\end{equation}
A lower ATC indicates better efficiency, signifying that the agent reaches clarity more quickly. Note that in cases where no premises are clarified (i.e., $|\mathcal{R}| = 0$), ATC is considered undefined and treated as \texttt{NaN} (not included in averages).

\textbf{Efficiency-Adjusted Recall (EAR)} provides a unified measure that balances clarification completeness (MPR) with interaction efficiency (ATC), penalizing agents that achieve high recall through excessive dialogue turns. The formal definition is provided in Appendix ~\ref{app:b}.

\section{Experiment}
\subsection{Experimental Setup}
\subsubsection{Models}
We evaluate eleven state-of-the-art code agents: Aider-GPT5\cite{Aider} (the Aider code editing framework powered by GPT-5), CodeX-GPT5 \cite{OpenAI}, GPT5-Coder \cite{hurst2024gpt,achiam2023gpt}, CodeLlama-70B \cite{roziere2023code}, Claude-Opus 4.1 \cite{Claude-a}, Claude-Sonnet 4.5 \cite{Claude-b}, O3, O4 \cite{OpenAI2025b}, DeepSeek-Coder \cite{guo2025deepseek}, Qwen2.5-Coder \cite{hui2024qwen2}, and Qwen3-Coder \cite{yang2025qwen3}. These models represent the current spectrum of commercial and open-source code agents, spanning diverse architectures (e.g., decoder-only, tool-augmented) and training paradigms (e.g., supervised fine-tuning, reinforcement learning).

\subsubsection{Evaluation Protocol}
Each model is evaluated on the ClarEval benchmark under both single-turn and multi-turn settings. In the single-turn setting, models receive only the ambiguous instruction and generate a single response containing clarification questions. In the multi-turn setting, we employ pre-defined dialogue scripts that simulate realistic user interactions, including partial answers and occasional introduction of new ambiguities. All tasks are derived from HumanEval via controlled injection of three ambiguity types: Missing Goal, Missing Premises, and Ambiguous Terminology.All experiments were conducted with temperature = 0.1 on NVIDIA A100 GPUs. Detailed hyperparameters and prompt templates are provided in Appendix ~\ref{app:a}.

\subsection{Overall Results}
Table ~\ref{tab:2} presents the overall performance across all ambiguity types, revealing significant variations in clarification capability among state-of-the-art code agents.
\subsubsection{Single-Turn Performance}
In the single-turn setting, GPT5-Coder achieves the highest KQC (0.867), demonstrating superior capability in inferring high-level user intents from ambiguous instructions. Qwen2.5-Coder (0.855) and Claude-Sonnet 4.5 (0.852) also perform strongly, though they lag behind GPT5-Coder by 8–9 percentage points.

For MPR, GPT5-Coder again leads (0.661), substantially outperforming the next-best models—Claude-Sonnet 4.5 (0.631) and Claude-Opus 4.1 (0.629)—by approximately 20 percentage points. This indicates that while several models can identify missing technical premises in isolation, only top-tier agents do so consistently across diverse ambiguity patterns.

\subsubsection{Multi-Turn Performance}
In the multi-turn setting, performance dynamics shift significantly. Claude-Opus 4.1 achieves the highest KQC (0.754), suggesting strong contextual understanding and effective follow-up questioning. For MPR, GPT5-Coder maintains its lead (0.951), closely followed by Claude-Opus 4.1 (0.930).

The ATC metric reveals a critical efficiency gap: GPT5-Coder achieves the lowest ATC (1.493), indicating it secures necessary clarifications in fewer dialogue turns. In stark contrast, Qwen2.5-Coder exhibits the highest ATC (2.396), requiring nearly 60\% more turns on average to clarify the same set of premises.

\subsection{Key Findings and Analysis}

\textbf{The Single-Turn vs. Multi-Turn Dichotomy:}

We observe a weak correlation between single-turn and multi-turn performance (Pearson r = 0.32, p < 0.05). For instance, Aider-GPT5 ranks 2nd in single-turn KQC (0.833) but drops to 10th in multi-turn KQC (0.376). Similarly, CodeLlama-70B shows a dramatic decline from single-turn (0.817) to multi-turn settings (0.427). This suggests that intent inference in isolation and dialogue-aware clarification are distinct skills, underscoring the necessity of evaluating both settings in a comprehensive benchmark.

\textbf{The Clarification Capability Spectrum:}

Our results reveal a clear spectrum of clarification competence. At the high end, GPT5-Coder and Claude-Opus 4.1 demonstrate balanced strength in intent understanding (KQC), technical detail recovery (MPR), and interaction efficiency (ATC). In the middle tier, models like DeepSeek-Coder and Claude-Sonnet 4.5 show competent but inconsistent performance. At the lower end, Aider-GPT5 and CodeLlama-70B struggle significantly in multi-turn settings, despite decent single-turn scores, revealing fundamental limitations in their interactive capabilities.

\textbf{Ambiguous Terminology as the Primary Challenge:}
As detailed in the fine-grained analysis, Ambiguous Terminology emerges as the most challenging ambiguity type across models and metrics. This suggests that while agents can readily detect missing information, they struggle to recognize when existing terms are vague or underspecified (e.g., "fast", "clean up", "user-friendly"). Given that such terminology is pervasive in real developer requests, this represents a key bottleneck for practical deployment.
Detailed breakdowns in Appendix Tables ~\ref{tab:10}-~\ref{tab:14} confirm this trend, showing lowest Pass@1 and highest ATC for terminology ambiguities across most models.

\begin{figure}[t]
    \centering
    \begin{tikzpicture}
        \begin{axis}[
            ybar,
            bar width=15pt,
            width=0.95\linewidth, %
            height=6cm,
            enlarge x limits=0.2,
            symbolic x coords={Terminology, Goal, Premises, Overall},
            xtick=data,
            ymin=0, ymax=100,
            ylabel={Pass@1 (\%)},
            legend style={at={(0.5,1.15)}, anchor=south, legend columns=-1, draw=none}, 
            nodes near coords, 
            nodes near coords style={font=\footnotesize, /pgf/number format/fixed, /pgf/number format/precision=1},
            grid=major,
            major grid style={dashed, gray!30}
        ]
            \addplot[fill=red!30, draw=red!80!black] coordinates {
                (Terminology, 6.71)
                (Goal, 9.76)
                (Premises, 10.37)
                (Overall, 8.94)
            };
            \addlegendentry{Baseline (Ambiguous)}

            \addplot[fill=green!30, draw=green!80!black] coordinates {
                (Terminology, 89.02)
                (Goal, 89.02)
                (Premises, 89.02)
                (Overall, 89.02)
            };
            \addlegendentry{Upper Bound (Clarified)}
            
        \end{axis}
    \end{tikzpicture}
    \caption{Performance Gap Analysis. Comparison of Pass@1 accuracy between the ambiguous baseline and the clarified upper bound across different ambiguity types. \textit{Ambiguous Terminology} proves to be the most challenging scenario for passive code generation.}
    \label{fig:pass_at_1_gap}
\end{figure}
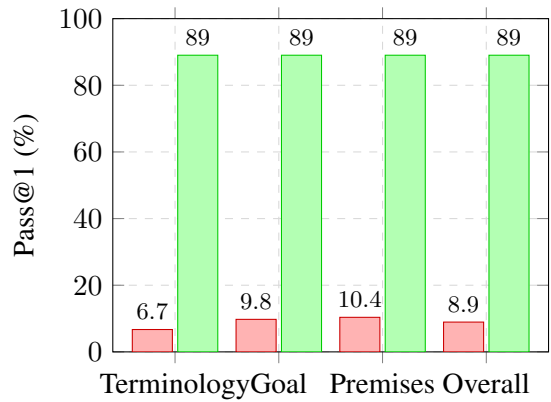

\section{Discussion}
\label{sec:discussion}

\subsection{The Impact of Clarification on Code Correctness}
\label{sec:impact_correctness}

To validate the practical utility of ClarEval, we conducted a controlled ablation study to quantify the impact of ambiguity on code generation correctness (Pass@1). We compared the performance of GPT-4o under two distinct conditions: an Ambiguous Baseline (generating code directly from underspecified instructions) and a Clarified Upper Bound (generating code after resolving ambiguities via oracle guidance). Note that for the baseline, we applied an oracle patch to fix function name mismatches, ensuring the scores reflect logic understanding rather than trivial interface misalignment.

\paragraph{The Catastrophic Ambiguity Gap.}
Figure~\ref{fig:pass_at_1_gap} illustrates a stark reality: despite the reasoning prowess of frontier models, ambiguity remains a catastrophic failure mode. While GPT-4o achieves a state-of-the-art Pass@1 of 89.02\% on clarified tasks, its performance plummets to 8.94\% under ambiguity. This substantial $\sim$80\% performance gap empirically demonstrates that without proactive clarification, even the most capable LLM-based agents fail to generate functionally correct code. Clarification is not merely an enhancement; it is a prerequisite for correctness. 

\paragraph{Semantic Vagueness as the Primary Bottleneck.}
Decomposing the results by ambiguity type reveals that Ambiguous Terminology ($A_T$) poses the most severe challenge, yielding the lowest Pass@1 of 6.71\% (compared to 9.76\% for Missing Goal and 10.37\% for Missing Premises). This indicates that semantic vagueness—where user intent is obscured by colloquialisms (e.g., ``clean up'', ``organize'')—introduces significantly higher variance in implementation logic than missing technical constraints.

\subsection{Beyond Detection: The Efficiency Frontier}
\label{sec:efficiency_frontier}

The high KQC scores achieved by top-tier models (e.g., GPT-5-Coder at 0.867) suggest that the community has largely solved the ``Ambiguity Detection'' problem for standard tasks. However, this does not render the benchmark obsolete; rather, it shifts the evaluation focus from competency to optimization.

The significant gaps exposed by ClarEval in the multi-turn setting and ATC metric demonstrate that current training paradigms prioritize correctness over collaboration costs. Effective code agents must evolve from ``Passive Generators'' that guess intent, to ``Active Partners'' that navigate ambiguity with minimal friction. 

\subsection{Implications for Agent Design}
\label{sec:implications}

The performance variations across ClarEval reveal critical directions for future agent development:

\begin{itemize}
    \item \textbf{Training for User-Centric Planning:} The poor efficiency of serial questioners points to a lack of user simulation in the training loop. Future training objectives should penalize excessive turns, encouraging models to anticipate multiple dependencies before yielding control to the user.
    
    \item \textbf{Semantic Grounding for Terminology:} The consistent failure on Ambiguous Terminology ($A_T$) suggests a disconnect between code reasoning and general semantic grounding. Agents require improved grounding in both technical domains and general world knowledge to bridge the gap between vague human intent and precise executable logic.
    
    \item \textbf{Adaptive Clarification Policies:} The observed ATC variations indicate that a one-size-fits-all questioning strategy is suboptimal. Future agents should develop adaptive policies that balance comprehensiveness (asking everything) and efficiency (asking only what is critical) based on the complexity of the context.
\end{itemize}

\section{Conclusion}

This work argues for a fundamental realignment in how we evaluate Code Agents: moving beyond the static ``passive generation'' paradigm toward active, efficiency-aware collaboration. We introduced ClarEval, not merely as a dataset, but as a diagnostic framework to measure this transition. By isolating communicative logic in a controlled environment and introducing efficiency-penalized metrics like ATC and EAR, we exposed a critical "Collaborative Gap" in current state-of-the-art models.

Our findings reveal that while modern agents have mastered the syntax of coding, they struggle with the semantics of partnership. High-performing coders often devolve into inefficient interrogators under ambiguity, burdening users with redundant dialogue. ClarEval provides the necessary granularity to detect and correct these interaction behaviors. Ultimately, we hope this benchmark serves as a catalyst for the community to prioritize Collaborative Quotient alongside code correctness, paving the way for agents that are not just powerful tools, but intuitive partners in the software engineering lifecycle.

\section*{Limitations}

While ClarEval provides comprehensive coverage of common ambiguity types, real-world scenarios may involve more complex or composite forms of ambiguity. Future work could expand the benchmark to include cross-cutting ambiguities and domain-specific underspecification. Additionally, exploring how clarification capability correlates with ultimate task success in real development environments would further validate the practical significance of our metrics.



\clearpage
\bibliography{custom}

@article{chen2021evaluating,
  title={Evaluating large language models trained on code},
  author={Chen, Mark and Tworek, Jerry and Jun, Heewoo and Yuan, Qiming and Pinto, Henrique Ponde De Oliveira and Kaplan, Jared and Edwards, Harri and Burda, Yuri and Joseph, Nicholas and Brockman, Greg and others},
  journal={arXiv preprint arXiv:2107.03374},
  year={2021}
}

@article{austin2021program,
  title={Program synthesis with large language models},
  author={Austin, Jacob and Odena, Augustus and Nye, Maxwell and Bosma, Maarten and Michalewski, Henryk and Dohan, David and Jiang, Ellen and Cai, Carrie and Terry, Michael and Le, Quoc and others},
  journal={arXiv preprint arXiv:2108.07732},
  year={2021}
}

@article{liu2023your,
  title={Is your code generated by chatgpt really correct? rigorous evaluation of large language models for code generation},
  author={Liu, Jiawei and Xia, Chunqiu Steven and Wang, Yuyao and Zhang, Lingming},
  journal={Advances in Neural Information Processing Systems},
  volume={36},
  pages={21558--21572},
  year={2023}
}

@article{zhang2024humaneval,
  title={HumanEval-V: Evaluating Visual Understanding and Reasoning Abilities of Large Multimodal Models Through Coding Tasks},
  author={Zhang, Fengji and Wu, Linquan and Bai, Huiyu and Lin, Guancheng and Li, Xiao and Yu, Xiao and Wang, Yue and Chen, Bei and Keung, Jacky},
  journal={arXiv preprint arXiv:2410.12381},
  year={2024}
}

@article{schick2023toolformer,
  title={Toolformer: Language models can teach themselves to use tools},
  author={Schick, Timo and Dwivedi-Yu, Jane and Dess{\`\i}, Roberto and Raileanu, Roberta and Lomeli, Maria and Hambro, Eric and Zettlemoyer, Luke and Cancedda, Nicola and Scialom, Thomas},
  journal={Advances in Neural Information Processing Systems},
  volume={36},
  pages={68539--68551},
  year={2023}
}

@article{yang2024swe,
  title={Swe-agent: Agent-computer interfaces enable automated software engineering},
  author={Yang, John and Jimenez, Carlos E and Wettig, Alexander and Lieret, Kilian and Yao, Shunyu and Narasimhan, Karthik and Press, Ofir},
  journal={Advances in Neural Information Processing Systems},
  volume={37},
  pages={50528--50652},
  year={2024}
}

@article{zhang2024codeagent,
  title={Codeagent: Enhancing code generation with tool-integrated agent systems for real-world repo-level coding challenges},
  author={Zhang, Kechi and Li, Jia and Li, Ge and Shi, Xianjie and Jin, Zhi},
  journal={arXiv preprint arXiv:2401.07339},
  year={2024}
}

@article{wu2024repomastereval,
  title={Repomastereval: Evaluating code completion via real-world repositories},
  author={Wu, Qinyun and Peng, Chao and Gao, Pengfei and Hu, Ruida and Gan, Haoyu and Jiang, Bo and Tang, Jinhe and Deng, Zhiwen and Guan, Zhanming and Gao, Cuiyun and others},
  journal={arXiv preprint arXiv:2408.03519},
  year={2024}
}

@inproceedings{puvvadi2025coding,
  title={Coding agents: A comprehensive survey of automated bug fixing systems and benchmarks},
  author={Puvvadi, Meghana and Arava, Sai Kumar and Santoria, Adarsh and Chennupati, Sesha Sai Prasanna and Puvvadi, Harsha Vardhan},
  booktitle={2025 IEEE 14th International Conference on Communication Systems and Network Technologies (CSNT)},
  pages={680--686},
  year={2025},
  organization={IEEE}
}

@article{chen2025acebench,
  title={ACEBench: Who Wins the Match Point in Tool Usage?},
  author={Chen, Chen and Hao, Xinlong and Liu, Weiwen and Huang, Xu and Zeng, Xingshan and Yu, Shuai and Li, Dexun and Wang, Shuai and Gan, Weinan and Huang, Yuefeng and others},
  journal={arXiv preprint arXiv:2501.12851},
  year={2025}
}

@article{Aider,
  title={Aider: Ai-assisted coding in your terminal with gpt. },
  author={Paul Gauthier.},
  journal={https://aider.chat/,2023. Accessed: 2025-05-15.},
  year={2025}
}

@article{OpenAI,
  title={Openai codex cli: Lightweight coding agent that runs in your terminal.},
  author={OpenAI.},
  journal={https://github.com/openai/codex, 2025.},
  year={2025}
}

@article{roziere2023code,
  title={Code llama: Open foundation models for code},
  author={Roziere, Baptiste and Gehring, Jonas and Gloeckle, Fabian and Sootla, Sten and Gat, Itai and Tan, Xiaoqing Ellen and Adi, Yossi and Liu, Jingyu and Sauvestre, Romain and Remez, Tal and others},
  journal={arXiv preprint arXiv:2308.12950},
  year={2023}
}

@article{OpenAI2025b,
  title={Introducing openai o3 and o4-mini.},
  author={OpenAI},
journal={https://openai.com/zh-Hans-CN/index/introducing-o3-and-o4-mini/},
  year={2025},
  
}

@article{hurst2024gpt,
  title={Introducing GPT-5},
  author={OpenAI},
  journal={https://openai.com/zh-Hans-CN/index/introducing-gpt-5/},
  year={2025}
}

@article{mcnall2024humanevalcomm,
  title={HumanEvalComm: Benchmarking the Communication Competence of Code Generation},
  author={McNall, Kyle and Qiao, Yiyun and others},
  journal={arXiv preprint arXiv:2406.00215},
  year={2024}
}

@article{Claude-a,
  title={Claude Opus 4.1},
  author={Anthropic.},
  journal={https://www.anthropic.com/news/claude-opus-4-1.},
  year={2025}
}

@article{Claude-b,
  title={Claude Sonnet 4.},
  author={Anthropic.},
  journal={Available online: https://www.anthropic.com/claude/opus.},
  year={2025}
}

@article{achiam2023gpt,
  title={Gpt-4 technical report},
  author={Achiam, Josh and Adler, Steven and Agarwal, Sandhini and Ahmad, Lama and Akkaya, Ilge and Aleman, Florencia Leoni and Almeida, Diogo and Altenschmidt, Janko and Altman, Sam and Anadkat, Shyamal and others},
  journal={arXiv preprint arXiv:2303.08774},
  year={2023}
}

@article{guo2025deepseek,
  title={Deepseek-r1: Incentivizing reasoning capability in llms via reinforcement learning},
  author={Guo, Daya and Yang, Dejian and Zhang, Haowei and Song, Junxiao and Zhang, Ruoyu and Xu, Runxin and Zhu, Qihao and Ma, Shirong and Wang, Peiyi and Bi, Xiao and others},
  journal={arXiv preprint arXiv:2501.12948},
  year={2025}
}

@article{yang2025qwen3,
  title={Qwen3 technical report},
  author={Yang, An and Li, Anfeng and Yang, Baosong and Zhang, Beichen and Hui, Binyuan and Zheng, Bo and Yu, Bowen and Gao, Chang and Huang, Chengen and Lv, Chenxu and others},
  journal={arXiv preprint arXiv:2505.09388},
  year={2025}
}

@article{hui2024qwen2,
  title={Qwen2. 5-coder technical report},
  author={Hui, Binyuan and Yang, Jian and Cui, Zeyu and Yang, Jiaxi and Liu, Dayiheng and Zhang, Lei and Liu, Tianyu and Zhang, Jiajun and Yu, Bowen and Lu, Keming and others},
  journal={arXiv preprint arXiv:2409.12186},
  year={2024}
}

@article{jain2024livecodebench,
  title={Livecodebench: Holistic and contamination free evaluation of large language models for code},
  author={Jain, Naman and Han, King and Gu, Alex and Li, Wen-Ding and Yan, Fanjia and Zhang, Tianjun and Wang, Sida and Solar-Lezama, Armando and Sen, Koushik and Stoica, Ion},
  journal={arXiv preprint arXiv:2403.07974},
  year={2024}
}

@article{qiu2025locobench,
  title={LoCoBench-Agent: An Interactive Benchmark for LLM Agents in Long-Context Software Engineering},
  author={Qiu, Jielin and Liu, Zuxin and Liu, Zhiwei and Murthy, Rithesh and Zhang, Jianguo and Chen, Haolin and Wang, Shiyu and Zhu, Ming and Yang, Liangwei and Tan, Juntao and others},
  journal={arXiv preprint arXiv:2511.13998},
  year={2025}
}

@article{wang2025codeflowbench,
  title={CodeFlowBench: A Multi-turn, Iterative Benchmark for Complex Code Generation},
  author={Wang, Sizhe and Wang, Zhengren and Ma, Dongsheng and Yu, Yongan and Ling, Rui and Li, Zhiyu and Xiong, Feiyu and Zhang, Wentao},
  journal={arXiv preprint arXiv:2504.21751},
  year={2025}
}

@article{zhan2025sr,
  title={SR-Eval: Evaluating LLMs on Code Generation under Stepwise Requirement Refinement},
  author={Zhan, Zexun and Gao, Shuzheng and Hu, Ruida and Gao, Cuiyun},
  journal={arXiv preprint arXiv:2509.18808},
  year={2025}
}

@article{zhang2025swe,
  title={SWE-bench Goes Live!},
  author={Zhang, Linghao and He, Shilin and Zhang, Chaoyun and Kang, Yu and Li, Bowen and Xie, Chengxing and Wang, Junhao and Wang, Maoquan and Huang, Yufan and Fu, Shengyu and others},
  journal={arXiv preprint arXiv:2505.23419},
  year={2025}
}

@article{wu2025humanevalcomm,
  title={Humanevalcomm: Benchmarking the communication competence of code generation for llms and llm agents},
  author={Wu, Jie JW and Fard, Fatemeh H},
  journal={ACM Transactions on Software Engineering and Methodology},
  volume={34},
  number={7},
  pages={1--42},
  year={2025},
  publisher={ACM New York, NY}
}

@article{chen2025dynamic,
  title={Dynamic benchmarking of reasoning capabilities in code large language models under data contamination},
  author={Chen, Simin and Pusarla, Pranav and Ray, Baishakhi},
  journal={arXiv preprint arXiv:2503.04149},
  year={2025}
}

@article{guan2025your,
  title={Is your benchmark (still) useful? dynamic benchmarking for code language models},
  author={Guan, Batu and Wu, Xiao and Yuan, Yuanyuan and Li, Shaohua},
  journal={arXiv preprint arXiv:2503.06643},
  year={2025}
}

@article{chen2024ppm,
  title={Ppm: Automated generation of diverse programming problems for benchmarking code generation models},
  author={Chen, Simin and Feng, Xiaoning and Han, Xiaohong and Liu, Cong and Yang, Wei},
  journal={Proceedings of the ACM on Software Engineering},
  volume={1},
  number={FSE},
  pages={1194--1215},
  year={2024},
  publisher={ACM New York, NY, USA}
}

@article{liu2023repobench,
  title={Repobench: Benchmarking repository-level code auto-completion systems},
  author={Liu, Tianyang and Xu, Canwen and McAuley, Julian},
  journal={arXiv preprint arXiv:2306.03091},
  year={2023}
}

@article{zhuo2024bigcodebench,
  title={Bigcodebench: Benchmarking code generation with diverse function calls and complex instructions},
  author={Zhuo, Terry Yue and Vu, Minh Chien and Chim, Jenny and Hu, Han and Yu, Wenhao and Widyasari, Ratnadira and Yusuf, Imam Nur Bani and Zhan, Haolan and He, Junda and Paul, Indraneil and others},
  journal={arXiv preprint arXiv:2406.15877},
  year={2024}
}

@article{huang2023agentcoder,
  title={Agentcoder: Multi-agent-based code generation with iterative testing and optimisation},
  author={Huang, Dong and Zhang, Jie M and Luck, Michael and Bu, Qingwen and Qing, Yuhao and Cui, Heming},
  journal={arXiv preprint arXiv:2312.13010},
  year={2023}
}

@article{huang2024code,
  title={Da-code: Agent data science code generation benchmark for large language models},
  author={Huang, Yiming and Luo, Jianwen and Yu, Yan and Zhang, Yitong and Lei, Fangyu and Wei, Yifan and He, Shizhu and Huang, Lifu and Liu, Xiao and Zhao, Jun and others},
  journal={arXiv preprint arXiv:2410.07331},
  year={2024}
}

@article{kasneci2023chatgpt,
  title={ChatGPT for good? On opportunities and challenges of large language models for education},
  author={Kasneci, Enkelejda and Se{\ss}ler, Kathrin and K{\"u}chemann, Stefan and Bannert, Maria and Dementieva, Daryna and Fischer, Frank and Gasser, Urs and Groh, Georg and G{\"u}nnemann, Stephan and H{\"u}llermeier, Eyke and others},
  journal={Learning and individual differences},
  volume={103},
  pages={102274},
  year={2023},
  publisher={Elsevier}
}

@article{chang2024survey,
  title={A survey on evaluation of large language models},
  author={Chang, Yupeng and Wang, Xu and Wang, Jindong and Wu, Yuan and Yang, Linyi and Zhu, Kaijie and Chen, Hao and Yi, Xiaoyuan and Wang, Cunxiang and Wang, Yidong and others},
  journal={ACM transactions on intelligent systems and technology},
  volume={15},
  number={3},
  pages={1--45},
  year={2024},
  publisher={ACM New York, NY}
}

@article{thirunavukarasu2023large,
  title={Large language models in medicine},
  author={Thirunavukarasu, Arun James and Ting, Darren Shu Jeng and Elangovan, Kabilan and Gutierrez, Laura and Tan, Ting Fang and Ting, Daniel Shu Wei},
  journal={Nature medicine},
  volume={29},
  number={8},
  pages={1930--1940},
  year={2023},
  publisher={Nature Publishing Group US New York}
}

@article{pan2024codev,
  title={Codev-Bench: How Do LLMs Understand Developer-Centric Code Completion?},
  author={Pan, Zhenyu and Cao, Rongyu and Cao, Yongchang and Ma, Yingwei and Li, Binhua and Huang, Fei and Liu, Han and Li, Yongbin},
  journal={arXiv preprint arXiv:2410.01353},
  year={2024}
}

\clearpage
\appendix

\section{Prompts and Implementation Details}
\label{app:a}

To ensure the reproducibility of ClarEval, we provide the detailed prompts used in our construction pipeline and evaluation framework. These details are extracted directly from our data synthesis engine and evaluation scripts.

\subsubsection{Implementation Details}

All experiments use standardized prompting strategies tailored to each model's interface, with a unified instruction encouraging proactive clarification. We set temperature=0.1 for deterministic generation and limit output length to 512 tokens. Evaluation is fully automated via the ClarEval framework, which parses model responses and computes metrics against human-annotated gold standards. All experiments were conducted on a cluster of NVIDIA A100 80GB GPUs.

\subsection{Ambiguity Injection Prompts}
\label{sec:injection_prompts_details}

We employed GPT-4o to transform standard HumanEval tasks into ambiguous variants. Table \ref{tab:injection_prompts} presents the specific prompts used for each ambiguity type.

\subsection{Dialogue Script Construction}

To enable multi-turn evaluation, we generated a ``User Simulator Script'' using the prompt in Table \ref{tab:script_prompt}.

\subsection{Evaluation Implementation}

In our evaluation, all agents shared a unified system instruction (Table \ref{tab:eval_prompts}). The User Simulator operated deterministically based on the generated scripts.

\section{Efficiency-Adjusted Recall (EAR)}
\label{app:b}
\subsection{Metric Definition}
\label{sec:ear_metric}

While \textit{Missing Premises Recall} (MPR) and \textit{Average Turns to Clarify} (ATC) evaluate clarification completeness and efficiency separately, a unified metric is desirable to penalize agents that achieve high recall only through excessive dialogue turns. Drawing inspiration from Discounted Cumulative Gain (DCG) in information retrieval, we introduce Efficiency-Adjusted Recall (EAR).

EAR measures the proportion of missing premises successfully clarified, with each clarified premise discounted by a logarithmic factor of the turn number in which it was resolved. Formally, let $\mathcal{P}$ be the set of all missing premises for a task, and $\mathcal{R} \subseteq \mathcal{P}$ be the set of clarified premises. For each clarified premise $p \in \mathcal{R}$, let $Turn(p)$ denote the dialogue turn index (starting from 1) where the premise was first resolved. The EAR score is defined as:

\begin{equation}
    EAR = \frac{1}{|\mathcal{P}|} \sum_{p \in \mathcal{R}} \frac{1}{\log_2(Turn(p) + 1)}
\end{equation}

Where:
\begin{itemize}
    \item A premise clarified in the \textbf{1st turn} contributes $\frac{1}{\log_2(1+1)} = 1.0$ (full value).
    \item A premise clarified in the \textbf{2nd turn} contributes $\frac{1}{\log_2(3)} \approx 0.63$.
    \item A premise clarified in the \textbf{3rd turn} contributes $\frac{1}{\log_2(4)} = 0.5$.
    \item Unclarified premises ($p \notin \mathcal{R}$) contribute $0$.
\end{itemize}

\textbf{Interpretation:} EAR creates a soft penalty for delayed clarification. An agent that recovers all information immediately ($EAR=1.0$) is superior to one that recovers all information over multiple turns ($EAR < 1.0$), which is in turn superior to an agent that fails to recover information ($EAR \approx 0$). This metric effectively addresses the trade-off between the ``Lazy Agent'' bias (low ATC but low MPR) and the ``Inefficient Agent'' bias (high MPR but high ATC).

\subsection{Quantitative Analysis}
\label{sec:appendix_ear}

\begin{table}[h]
    \centering
    \small
    \begin{tabular}{lcccc}
    \toprule
    \textbf{Model} & \textbf{MPR} & \textbf{ATC} & \textbf{EAR (Approx)} & \textbf{Rank} \\
    \midrule
    \textbf{GPT5-Coder} & \textbf{0.951} & \textbf{1.493} & \textbf{0.722} & 1 \\
    Claude-Opus 4.1 & 0.930 & 1.900 & 0.605 & 2 \\
    Claude-Sonnet 4.5 & 0.823 & 1.719 & 0.570 & 3 \\
    DeepSeek-Coder & 0.796 & 1.772 & 0.541 & 4 \\
    O3 & 0.791 & 1.771 & 0.538 & 5 \\
    CodeX-GPT5 & 0.748 & 1.681 & 0.526 & 6 \\
    O4 & 0.712 & 1.576 & 0.522 & 7 \\
    Qwen3-Coder & 0.640 & 1.645 & 0.456 & 8 \\
    Qwen2.5-Coder & 0.643 & 2.396 & 0.365 & 9 \\
    Aider-GPT5 & 0.396 & 1.576 & 0.290 & 10 \\
    CodeLlama-70B & 0.385 & 1.602 & 0.279 & 11 \\
    \bottomrule
    \end{tabular}
    \caption{Efficiency-Adjusted Recall (EAR) rankings. Higher EAR indicates better balance between identifying missing premises and minimizing user dialogue turns.}
    \label{tab:ear_results}
\end{table}

Table \ref{tab:ear_results} presents the ranked EAR scores. This metric reveals significant performance distinctions that are less visible when viewing MPR and ATC in isolation:

\begin{itemize}
    \item \textbf{Dominance of GPT5-Coder:} GPT5-Coder achieves a dominant EAR of \textbf{0.722}, outperforming the second-best model (Claude-Opus 4.1) by a substantial margin ($\sim$19\%). This confirms that GPT5-Coder is not only thorough but also exceptionally efficient, often resolving ambiguities in fewer turns.
    
    \item \textbf{The Efficiency Penalty:} While Qwen2.5-Coder achieves a comparable MPR (0.643) to Qwen3-Coder (0.640), its EAR drops significantly (0.365 vs. 0.456). This penalty reflects Qwen2.5-Coder's high ATC (2.396), indicating a tendency to ask fragmented questions over many turns, which degrades the user experience.
\end{itemize}

\subsection{The Efficiency-Cognitive Load Trade-off}
\label{sec:efficiency_case_study}

A critical finding of our study is the mismatch between capability and interaction efficiency in modern code agents. While open-weights models like Qwen2.5-Coder achieve competitive intent detection scores (KQC: 0.855), their interaction efficiency lags significantly behind frontier models like GPT-5-Coder (ATC: 2.396 vs. 1.493). We interpret this not merely as a performance gap, but as a trade-off in clarification planning.

\paragraph{Batching vs. Serial Strategies.}
Our qualitative audit reveals distinct behavioral patterns:
\begin{itemize}
    \item \textbf{The Batching Strategy (GPT-5-Coder):} Frontier models tend to employ a ``Parallel'' approach. When faced with composite ambiguities (e.g., missing both sort order and data type), these agents consolidate inquiries into a single, structured turn. While this increases the per-turn reading load for the user, it significantly minimizes interaction latency (Low ATC) and accelerates the time-to-solution.
    \item \textbf{The Fragmented Strategy (Qwen2.5-Coder):} Conversely, other models often exhibit a ``Serial Chain'' behavior. While a serial approach can theoretically reduce per-turn cognitive load, our analysis shows that models like Qwen2.5 do not employ this strategically. Instead, they exhibit \textit{fragmented reasoning}—asking about a data type in Turn 1, waiting for feedback, and only then realizing they also need the sort order in Turn 3.
\end{itemize}

This distinction is crucial: whereas strategic serial questioning can be a valid user-centric choice, the observed fragmented questioning indicates a failure in global planning. The high ATC in these models stems from an inability to anticipate dependencies, imposing a severe ``Cognitive Tax'' of friction on the user. Thus, ClarEval's ATC metric serves as a proxy for an agent's ability to plan clarifications efficiently.

\section{Dialogue Script Construction (Multi-Turn)}
\label{app:c}
To evaluate the agent's ability to recover this missing information, we constructed a ``User Simulator Script'' $\mathcal{S}_i$ for each task. The script $\mathcal{S}_i$ is derived from the original ground truth $T_i$ and acts as a conditional lookup table for the simulator.

Each script is structured as a set of trigger-response pairs:
$$ \mathcal{S}_i = \{ (I_{trigger}, R_{content}) \} $$
where $I_{trigger}$ represents the intent the agent asks about (e.g., ``Input Format'', ``Edge Case''), and $R_{content}$ is the pre-defined user answer restored from the ground truth.

For example, in a task where the \textit{Missing Premises} strategy removed the constraint ``The list must be sorted descending'', the script records this as a conditional response: \textit{If Agent asks about order $\rightarrow$ Reply ``Sort it descending.''} This ensures that the evaluation of multi-turn capabilities is deterministic and consistent across all models.

\section{Robustness Validation of the User Simulator}
\label{sec:appendix_e}

To address concerns regarding the potential rigidity of the rule-based matching mechanism used in our User Simulator, we conducted a rigorous validation experiment. Our goal was to determine if the rule-based keywords (Triggers) sufficiently capture the semantic intent of clarification questions generated by advanced agents.

\subsection{Methodology: Semantic Consistency Check}
We randomly sampled 200 interaction turns from the evaluation logs of the top-performing models (GPT-5 Coder and Claude-Opus 4.1). We then employed a ``LLM-as-a-Judge'' approach using GPT-4o to semantically evaluate the agent's questions against the ground truth intent, independent of the keyword triggers.

The LLM Judge was instructed to classify each agent question into one of two categories:
\begin{itemize}
    \item \textbf{Hit (Semantic):} The question effectively asks for the missing information required by the ground truth.
    \item \textbf{Miss (Semantic):} The question is irrelevant, too vague, or hallucinates constraints.
\end{itemize}

We compared these semantic labels with the deterministic labels (Hit/Miss) generated by our rule-based User Simulator.

\subsection{Results}
Table 8 presents the confusion matrix between the Rule-Based Simulator and the LLM Semantic Judge.

The analysis yields the following insights:
\begin{itemize}
    \item \textbf{Agreement Rate:} The overall agreement between our rule-based script and the semantic judge is \textbf{96.5\%}.
    \item \textbf{False Negatives (2.5\%):} In only 5 out of 200 cases did the rule-based simulator fail to recognize a valid question. This typically occurred when agents used highly metaphorical language (e.g., asking about ``data hygiene'' instead of ``null values'').
    \item \textbf{Conclusion:} The high alignment confirms that our trigger keyword sets are sufficiently diverse and robust. The rule-based approach provides a computationally efficient and deterministic evaluation without significantly penalizing the semantic flexibility of SOTA models.
\end{itemize}

\section{Qualitative Analysis of Ambiguity Resolution}
\label{sec:appendix_f}

To provide deeper insights into why models struggle significantly with \textbf{Ambiguous Terminology ($A_T$)}, we conducted a qualitative analysis of failure cases. We identified three primary error modes that hinder clarification performance.

\subsection{Taxonomy of Clarification Failures}
\begin{itemize}
    \item \textbf{Assumptive Generation (The ``Silent Failure''):} The agent ignores the ambiguity entirely and generates code based on a high-probability assumption (e.g., assuming ``clean up'' always means dropna. This is the most common failure mode in single-turn settings.
    \item \textbf{Generic Querying:} The agent recognizes something is wrong but asks a non-actionable question, such as \textit{``Is there anything else you need?''} or \textit{``Can you provide more details?''} without specifying \textit{what} details are missing. These do not trigger the User Simulator's specific responses.
    \item \textbf{Hallucinated Constraints:} Instead of clarifying, the agent invents its own constraints (e.g., \textit{``I will assume the data is sorted''}), effectively bypassing the clarification process.
\end{itemize}

\subsection{Case Study: ``Ambiguous Terminology''}
Table 9 demonstrates a concrete example of how different agents handle the ambiguous term ``Organize'' in the task: \textit{``Organize the list of items based on the 'id'.''}

\subsection{Implications}
This analysis suggests that future model training should focus on \textbf{Semantic Sensitivity}—teaching models to flag non-technical descriptors (like ``fast'', ``best'', ``organize'') as triggers for mandatory clarification, rather than treating them as stylistic noise.

\section{Fine-Grained Analysis by Ambiguity Type}
\subsection{KQC Breakdown}

Appendix Table 10 and Table 11 shows KQC performance across different ambiguity types. In single-turn settings, all models perform best on Missing Goal scenarios, with GPT5-Coder achieving 0.945. Performance on Ambiguous Terminology is consistently lowest across models (e.g., GPT5-Coder: 0.918; Claude-Opus: 0.817), indicating this is the most challenging ambiguity type for intent inference.

In multi-turn settings, most models suffer significant KQC degradation. Notably, Aider-GPT5 drops from 0.857 to 0.364 on Missing Goal, and CodeLlama-70B falls from 0.854 to 0.434, revealing severe limitations in context management and iterative intent refinement.

\subsection{MPR Breakdown}

As shown in Appendix Table 12 and Table 13, GPT5-Coder consistently achieves high MPR across all ambiguity types, particularly excelling in multi-turn Missing Premises scenarios (0.943). Claude-Opus 4.1 also shows robust multi-turn performance (0.930–0.935 across types), indicating reliable technical clarification capabilities.

Most models show improved MPR in multi-turn vs. single-turn settings, validating the value of extended interaction for uncovering implementation details. However, Aider-GPT5 and CodeLlama-70B show minimal improvement, suggesting their multi-turn reasoning mechanisms are underdeveloped.

\subsection{Efficiency Analysis}

Appendix Table 14 reveals that GPT5-Coder achieves the most consistent efficiency across ambiguity types (ATC: 1.422–1.602), often asking multiple critical questions in a single turn. Conversely, Qwen2.5-Coder shows the highest ATC values (2.338–2.486), reflecting a sequential, less efficient questioning strategy. Interestingly, ATC patterns are relatively stable across ambiguity types for most models, suggesting that questioning efficiency is a model-intrinsic trait, independent of ambiguity nature.

\section{The Value of ClarEval}

ClarEval successfully moves beyond functional correctness, providing a standardized framework to assess and, ultimately, improve the collaborative intelligence of code agents. By quantifying capabilities across intent inference, technical clarification, and interaction efficiency, our benchmark enables targeted improvements in agent design. This shift from correctness-only to collaboration-aware evaluation is essential for developing the next generation of usable, trustworthy, and human-aligned programming assistants.

\begin{table*}[t]
\centering
\small
\renewcommand{\arraystretch}{1.4} %
\begin{tabular}{p{0.15\textwidth} p{0.40\textwidth} p{0.38\textwidth}}
\toprule
\textbf{Ambiguity Type} & \textbf{Injection Prompt Strategy} & \textbf{Transformation Example} \\
\midrule

\textbf{Missing Goal} \newline ($A_G$) & 
\textbf{Strategy: Logic Omission} \newline
The prompt simulates a user who provides data context but forgets the objective.
\begin{itemize}
    \item Identify core algorithmic goal ("what to do").
    \item Remove the specific logic entirely.
    \item Retain input context descriptions.
\end{itemize} & 
\textit{Original:} "Given a list of integers, \underline{return the sum of all even numbers}." \newline
\centerline{$\downarrow$} 
\textit{Ambiguous:} "I have a list of integers. Write a function to \underline{process them}." \\
\midrule

\textbf{Missing Premises} \newline ($A_P$) & 
\textbf{Strategy: Constraint Removal} \newline
The prompt simulates a user who has a goal but omits critical boundary conditions.
\begin{itemize}
    \item Keep the high-level goal.
    \item Identify and remove specific constraints (e.g., "shortest", "starts with").
\end{itemize} & 
\textit{Original:} "Find the \underline{shortest} palindrome that \underline{starts with} the given string." \newline
\centerline{$\downarrow$} 
\textit{Ambiguous:} "Write a function that \underline{finds a palindrome} using the given string." \\
\midrule

\textbf{Ambiguous Terms} \newline ($A_T$) & 
\textbf{Strategy: Vague Replacement} \newline
The prompt simulates a non-technical user employing colloquial language.
\begin{itemize}
    \item Identify precise technical terms (e.g., "sort", "dict").
    \item Replace with vague synonyms (e.g., "organize", "structure") without removing context.
\end{itemize} & 
\textit{Original:} "\underline{Sort} the list of \underline{dictionaries} by the 'id' key." \newline
\centerline{$\downarrow$} 
\textit{Ambiguous:} "\underline{Organize} the list of \underline{items} based on the 'id'." \\

\bottomrule
\end{tabular}
\caption{The Ambiguity Injection strategies used to construct ClarEval. We apply distinct transformation rules—Omission ($A_G$), Constraint Removal ($A_P$), and Vague Replacement ($A_T$)—to systematically generate diverse types of underspecification from the original HumanEval tasks.}
\label{tab:prompt_strategies}
\end{table*}

\begin{table*}[h]
    \centering
    \small
    \renewcommand{\arraystretch}{1.3}
    \begin{tabularx}{\textwidth}{lX}
        \toprule
        \textbf{Type} & \textbf{Injection Prompt Template} \\
        \midrule
        \textbf{System} & You are a software requirement engineer. Your task is to rewrite clear coding problems into ambiguous or underspecified versions to test an AI assistant's clarification capabilities. \\
        \midrule
        \textbf{Missing Goal} & \textbf{Task:} Rewrite the following problem description to introduce ambiguity by \textbf{removing the explicit goal}. \newline
        \textbf{Instructions:} \newline
        1. Keep the context (input data description, setup). \newline
        2. \textbf{Remove} the specific instruction of what to calculate or return. \newline
        3. Replace the goal with a vague request (e.g., "process the data", "handle the input"). \newline
        4. Ensure the code cannot be written correctly without asking "What should I do with this data?". \\
        \midrule
        \textbf{Missing Premises} & \textbf{Task:} Rewrite the following problem description to introduce ambiguity by \textbf{omitting necessary constraints}. \newline
        \textbf{Instructions:} \newline
        1. Keep the high-level goal clear (e.g., "sort the list"). \newline
        2. \textbf{Remove} specific constraints required for a unique solution (e.g., sort order, handling of duplicates, edge cases like empty lists). \newline
        3. The resulting description should look plausible but lead to multiple possible valid implementations. \\
        \midrule
        \textbf{Ambiguous Terms} & \textbf{Task:} Rewrite the following problem description to introduce ambiguity by \textbf{using vague terminology}. \newline
        \textbf{Instructions:} \newline
        1. Identify precise technical terms in the original description (e.g., "dictionary", "prime number", "recursion"). \newline
        2. \textbf{Replace} them with non-technical, colloquial, or ambiguous synonyms (e.g., "lookup list", "special number", "repeated logic"). \newline
        3. Do not remove the logic, but make the \textit{language} imprecise. \\
        \bottomrule
    \end{tabularx}
    \caption{The prompt templates used to inject three types of ambiguity into coding tasks.}
    \label{tab:injection_prompts}
\end{table*}

\begin{table*}[h]
    \centering
    \small
    \renewcommand{\arraystretch}{1.3}
    \begin{tabularx}{\textwidth}{lX}
        \toprule
        \textbf{Component} & \textbf{Content} \\
        \midrule
        \textbf{Script Gen. Prompt} & \textbf{Task:} Compare the Original Task and the Ambiguous Task below. Identify the missing information that an agent would need to ask to recover the Original Task's requirements. \newline
        \textbf{Input:} Original Task: \{original\_prompt\} | Ambiguous Task: \{ambiguous\_prompt\} \newline
        \textbf{Output Format (JSON):} Generate a list of Key-Value pairs where: \newline
        - \textbf{Trigger:} A short keyword or phrase representing the missing intent (e.g., "Sort Order"). \newline
        - \textbf{Response:} The specific detail from the Original Task (e.g., "Sort in descending order"). \\
        \bottomrule
    \end{tabularx}
    \caption{The prompt used to generate the User Simulator's lookup table (Ground Truth intents).}
    \label{tab:script_prompt}
\end{table*}

\begin{table*}[h]
    \centering
    \small
    \renewcommand{\arraystretch}{1.3}
    \begin{tabularx}{\textwidth}{lX}
        \toprule
        \textbf{Component} & \textbf{Implementation Logic / Prompt} \\
        \midrule
        \textbf{Agent System Prompt} & You are a helpful and professional coding assistant. When you receive a coding task from the user, please follow these steps: \newline
        1. Carefully analyze the user's instruction. \newline
        2. If the instruction is \textbf{ambiguous, underspecified, or lacks critical details}, you should \textbf{ask clarifying questions} to the user to resolve the ambiguity. \newline
        3. Do not guess the user's intent if it is unclear. \newline
        4. Once you have sufficient information, generate the code to solve the problem. \\
        \midrule
        \textbf{User Simulator} & \textbf{Mechanism:} Deterministic, Rule-Based Simulation using Pre-defined Scripts. \newline
        \textbf{Role Setup:} The simulator follows a pre-defined script $S_i$ containing trigger-response pairs derived from the ground truth $T_{clear}$. \newline
        \textbf{Interaction Logic:} \newline
        1. \textbf{Input Matching:} The agent's question is matched against pre-defined trigger keywords/phrases in $S_i$. \newline
        2. \textbf{Response Retrieval:} \newline
        -- If a trigger match is found $\rightarrow$ Return the corresponding pre-written response from the script. \newline
        -- If no match is found $\rightarrow$ Return fallback: "I don't have specific requirements for that; please follow standard best practices." \newline
        \textbf{Termination:} Upon code generation or reaching the maximum number of turns ($T=5$). \\
        \bottomrule
    \end{tabularx}
    \caption{Details of the Agent System Prompt and the deterministic, rule-based User Simulator logic used in \texttt{eval.py}. The simulator operates based on pre-defined scripts to ensure reproducibility and fairness across all model evaluations.}
    \label{tab:eval_prompts}
\end{table*}

\begin{table*}[h]
\centering

\label{tab:confusion_matrix}
\begin{tabular}{l|cc}
\toprule
 & \textbf{LLM Judge: Hit} & \textbf{LLM Judge: Miss} \\
\midrule
\textbf{Rule-Based: Hit} & \textbf{84 (True Positive)} & 2 (False Positive) \\
\textbf{Rule-Based: Miss} & 5 (False Negative) & \textbf{109 (True Negative)} \\
\bottomrule
\end{tabular}
\caption{Confusion Matrix of Simulator Decisions vs. LLM Semantic Judge. The high agreement rate confirms the reliability of the rule-based mechanism.}
\end{table*}

\begin{table*}[t]
\centering
\small
\renewcommand{\arraystretch}{1.3}

\label{tab:case_study}
\begin{tabular}{p{0.15\linewidth} p{0.35\linewidth} p{0.15\linewidth} p{0.25\linewidth}}
\toprule
\textbf{Model} & \textbf{Response Strategy} & \textbf{Outcome} & \textbf{Analysis} \\
\midrule
\textbf{GPT-5 Coder} \newline (Strong) & \textbf{Clarification Question:} \newline \textit{``By 'organize', do you mean sorting the list? If so, should it be in ascending or descending order?''} & \textbf{Success} \newline (Hit) & The agent correctly mapped the vague term ``organize'' to the technical concept ``sort'' and identified the missing parameter (order). \\
\midrule
\textbf{Qwen2.5-Coder} \newline (Inefficient) & \textbf{Clarification Question:} \newline \textit{``What is the structure of the items?''} \newline \textit{(Next Turn): ``How should I handle the ID?''} & \textbf{Partial Success} \newline (High ATC) & The agent asked tangential questions first. It took 3 turns to finally ask about the sorting order, resulting in a high ATC score. \\
\midrule
\textbf{CodeLlama-70B} \newline (Weak) & \textbf{Direct Action:} \newline \textit{``Here is the python code to sort the list of dictionaries by ID in ascending order...''} & \textbf{Failure} \newline (Miss) & The model exhibited \textbf{Assumptive Generation}. It failed to recognize ``organize'' as ambiguous and arbitrarily chose ``ascending sort,'' leading to potential misalignment with user intent. \\
\bottomrule
\end{tabular}
\caption{Comparison of Agent Responses to Ambiguous Terminology. While GPT-5 Coder correctly identifies the need to clarify the sort order, weaker models either ask inefficient questions or make unsafe assumptions.}
\end{table*}


\begin{table*}[htbp]
\centering

\begin{tabular}{@{}lccc@{}}
\toprule
\textbf{Model} & \textbf{Missing Goal} & \textbf{Missing Premises} & \textbf{Ambiguous Terminology} \\
\midrule
Aider-GPT5      & \cellcolor{red-dark!23}0.600 & \cellcolor{orange-dark!15}0.593 & \cellcolor{yellow-dark!20}0.571 \\
CodeX-GPT5      & \cellcolor{red-dark!5}0.551 & \cellcolor{orange-dark!12}0.585 & \cellcolor{yellow-dark!8}0.536 \\
GPT5-Coder      & \cellcolor{red-dark!100}0.830 & \cellcolor{orange-dark!100}0.661 & \cellcolor{yellow-dark!100}0.810 \\
CodeLlama-70B   & \cellcolor{red-dark!1}0.539 & \cellcolor{orange-dark!0}0.555 & \cellcolor{yellow-dark!0}0.512 \\
Claude-opus4.1  & \cellcolor{red-dark!25}0.607 & \cellcolor{orange-dark!36}0.650 & \cellcolor{yellow-dark!40}0.631 \\
Claude-sonnet-4.5 & \cellcolor{red-dark!32}0.625 & \cellcolor{orange-dark!31}0.636 & \cellcolor{yellow-dark!41}0.633 \\
O3              & \cellcolor{red-dark!13}0.572 & \cellcolor{orange-dark!13}0.588 & \cellcolor{yellow-dark!15}0.556 \\
O4              & \cellcolor{red-dark!0}0.536 & \cellcolor{orange-dark!0}0.554 & \cellcolor{yellow-dark!10}0.542 \\
DeepSeek-Coder  & \cellcolor{red-dark!18}0.586 & \cellcolor{orange-dark!13}0.589 & \cellcolor{yellow-dark!26}0.588 \\
Qwen2.5-Coder   & \cellcolor{red-dark!28}0.615 & \cellcolor{orange-dark!24}0.618 & \cellcolor{yellow-dark!41}0.633 \\
Qwen3-Coder     & \cellcolor{red-dark!19}0.590 & \cellcolor{orange-dark!25}0.621 & \cellcolor{yellow-dark!18}0.565 \\
\bottomrule
\end{tabular}
\caption{Fine-grained results for KQC in the Single-Turn setting. We report the agent's performance across three distinct ambiguity types: Missing Goal, Missing Premises, and Ambiguous Terminology. (Note: Higher values indicate a stronger capability to infer the correct user intent immediately from the initial underspecified instruction.)}
\label{tab:10}
\end{table*}


\begin{table*}[htbp] 
\centering

\label{tab:11}
\begin{tabular}{@{}lccc@{}}
\toprule
\textbf{Model} & \textbf{Missing Goal} & \textbf{Missing Premises} & \textbf{Ambiguous Terminology} \\
\midrule
Aider-GPT5      & \cellcolor{red-dark!0}0.364 & \cellcolor{orange-dark!0}0.373 & \cellcolor{yellow-dark!0}0.392 \\
CodeX-GPT5      & \cellcolor{red-dark!71}0.639 & \cellcolor{orange-dark!63}0.623 & \cellcolor{yellow-dark!64}0.615 \\
GPT5-Coder      & \cellcolor{red-dark!36}0.505 & \cellcolor{orange-dark!46}0.554 & \cellcolor{yellow-dark!39}0.529 \\
CodeLlama-70B   & \cellcolor{red-dark!18}0.434 & \cellcolor{orange-dark!4}0.389 & \cellcolor{yellow-dark!19}0.459 \\
Claude-opus4.1  & \cellcolor{red-dark!100}0.753 & \cellcolor{orange-dark!100}0.768 & \cellcolor{yellow-dark!100}0.742 \\
Claude-sonnet-4.5 & \cellcolor{red-dark!65}0.618 & \cellcolor{orange-dark!70}0.649 & \cellcolor{yellow-dark!68}0.631 \\
O3              & \cellcolor{red-dark!53}0.569 & \cellcolor{orange-dark!54}0.586 & \cellcolor{yellow-dark!49}0.562 \\
O4              & \cellcolor{red-dark!44}0.535 & \cellcolor{orange-dark!48}0.562 & \cellcolor{yellow-dark!36}0.517 \\
DeepSeek-Coder  & \cellcolor{red-dark!74}0.651 & \cellcolor{orange-dark!83}0.702 & \cellcolor{yellow-dark!80}0.671 \\
Qwen2.5-Coder   & \cellcolor{red-dark!43}0.531 & \cellcolor{orange-dark!45}0.549 & \cellcolor{yellow-dark!46}0.551 \\
Qwen3-Coder     & \cellcolor{red-dark!82}0.681 & \cellcolor{orange-dark!78}0.683 & \cellcolor{yellow-dark!71}0.639 \\
\bottomrule
\end{tabular}
\caption{Fine-grained results for KQC in the Multi-Turn setting. We report the agent's performance across three distinct ambiguity types: Missing Goal, Missing Premises, and Ambiguous Terminology. (Note: Higher values indicate a stronger capability to infer the correct user intent immediately from the initial underspecified instruction.)}
\end{table*}


\begin{table*}[htbp]
\centering

\label{tab:12}
\begin{tabular}{@{}lccc@{}}
\toprule
\textbf{Model} & \textbf{Missing Goal} & \textbf{Missing Premises} & \textbf{Ambiguous Terminology} \\
\midrule
Aider-GPT5      & \cellcolor{red-dark!23}0.600 & \cellcolor{orange-dark!15}0.593 & \cellcolor{yellow-dark!20}0.571 \\
CodeX-GPT5      & \cellcolor{red-dark!5}0.551 & \cellcolor{orange-dark!12}0.585 & \cellcolor{yellow-dark!8}0.536 \\
GPT5-Coder      & \cellcolor{red-dark!100}0.830 & \cellcolor{orange-dark!100}0.661 & \cellcolor{yellow-dark!100}0.810 \\
CodeLlama-70B   & \cellcolor{red-dark!1}0.539 & \cellcolor{orange-dark!0}0.555 & \cellcolor{yellow-dark!0}0.512 \\
Claude-opus4.1  & \cellcolor{red-dark!25}0.607 & \cellcolor{orange-dark!36}0.650 & \cellcolor{yellow-dark!40}0.631 \\
Claude-sonnet-4.5 & \cellcolor{red-dark!32}0.625 & \cellcolor{orange-dark!31}0.636 & \cellcolor{yellow-dark!41}0.633 \\
O3              & \cellcolor{red-dark!13}0.572 & \cellcolor{orange-dark!13}0.588 & \cellcolor{yellow-dark!15}0.556 \\
O4              & \cellcolor{red-dark!0}0.536 & \cellcolor{orange-dark!0}0.554 & \cellcolor{yellow-dark!10}0.542 \\
DeepSeek-Coder  & \cellcolor{red-dark!18}0.586 & \cellcolor{orange-dark!13}0.589 & \cellcolor{yellow-dark!26}0.588 \\
Qwen2.5-Coder   & \cellcolor{red-dark!28}0.615 & \cellcolor{orange-dark!24}0.618 & \cellcolor{yellow-dark!41}0.633 \\
Qwen3-Coder     & \cellcolor{red-dark!19}0.590 & \cellcolor{orange-dark!25}0.621 & \cellcolor{yellow-dark!18}0.565 \\
\bottomrule
\end{tabular}
\caption{Fine-grained results for PIR in the Single-Turn setting. We report the agent's performance across three distinct ambiguity types: Missing Goal, Missing Premises, and Ambiguous Terminology. (Note: Higher values indicate a stronger capability to infer the correct user intent immediately from the initial underspecified instruction.)}
\end{table*}


\begin{table*}[htbp]
\centering

\label{tab:13}
\begin{tabular}{@{}lccc@{}}
\toprule
\textbf{Model} & \textbf{Missing Goal} & \textbf{Missing Premises} & \textbf{Ambiguous Terminology} \\
\midrule
Aider-GPT5      & \cellcolor{red-dark!4}0.393 & \cellcolor{orange-dark!5}0.389 & \cellcolor{yellow-dark!1}0.405 \\
CodeX-GPT5      & \cellcolor{red-dark!62}0.729 & \cellcolor{orange-dark!66}0.747 & \cellcolor{yellow-dark!66}0.768 \\
GPT5-Coder      & \cellcolor{red-dark!100}0.954 & \cellcolor{orange-dark!100}0.943 & \cellcolor{yellow-dark!100}0.956 \\
CodeLlama-70B   & \cellcolor{red-dark!0}0.367 & \cellcolor{orange-dark!0}0.360 & \cellcolor{yellow-dark!5}0.430 \\
Claude-opus4.1  & \cellcolor{red-dark!96}0.930 & \cellcolor{orange-dark!97}0.924 & \cellcolor{yellow-dark!96}0.867 \\
Claude-sonnet-4.5 & \cellcolor{red-dark!76}0.809 & \cellcolor{orange-dark!79}0.822 & \cellcolor{yellow-dark!78}0.837 \\
O3              & \cellcolor{red-dark!78}0.822 & \cellcolor{orange-dark!71}0.777 & \cellcolor{yellow-dark!67}0.773 \\
O4              & \cellcolor{red-dark!59}0.715 & \cellcolor{orange-dark!61}0.718 & \cellcolor{yellow-dark!54}0.703 \\
DeepSeek-Coder  & \cellcolor{red-dark!74}0.799 & \cellcolor{orange-dark!76}0.803 & \cellcolor{yellow-dark!69}0.785 \\
Qwen2.5-Coder   & \cellcolor{red-dark!49}0.657 & \cellcolor{orange-dark!45}0.623 & \cellcolor{yellow-dark!44}0.649 \\
Qwen3-Coder     & \cellcolor{red-dark!47}0.644 & \cellcolor{orange-dark!47}0.635 & \cellcolor{yellow-dark!43}0.641 \\
\bottomrule
\end{tabular}
\caption{Fine-grained results for MPR in the Multi-Turn setting. We report the agent's performance across three distinct ambiguity types: Missing Goal, Missing Premises, and Ambiguous Terminology. (Note: Higher values indicate a stronger capability to infer the correct user intent immediately from the initial underspecified instruction.)}
\end{table*}

\begin{table*}[htbp]
\centering

\begin{tabular}{lccc}
\toprule
\textbf{Model} & \textbf{Missing Goal} & \textbf{Missing Premises} & \textbf{Ambiguous Terminology} \\
\midrule
Aider-GPT5        & \cellcolor{red-dark!18}1.603 & \cellcolor{orange-dark!30}1.548 & \cellcolor{yellow-dark!22}1.577 \\
CodeX-GPT5        & \cellcolor{red-dark!16}1.625 & \cellcolor{orange-dark!0}1.766 & \cellcolor{yellow-dark!14}1.651 \\
GPT5-Coder        & \cellcolor{red-dark!100}1.422 & \cellcolor{orange-dark!100}1.455 & \cellcolor{yellow-dark!20}1.602 \\
CodeLlama-70B     & \cellcolor{red-dark!32}1.533 & \cellcolor{orange-dark!33}1.466 & \cellcolor{yellow-dark!0}1.807 \\
Claude-opus4.1    & \cellcolor{red-dark!0}1.898 & \cellcolor{orange-dark!1}1.869 & \cellcolor{yellow-dark!0}1.933 \\
Claude-sonnet-4.5 & \cellcolor{red-dark!31}1.539 & \cellcolor{orange-dark!0}1.885 & \cellcolor{yellow-dark!7}1.732 \\
O3                & \cellcolor{red-dark!1}1.840 & \cellcolor{orange-dark!9}1.701 & \cellcolor{yellow-dark!6}1.772 \\
O4                & \cellcolor{red-dark!42}1.458 & \cellcolor{orange-dark!16}1.629 & \cellcolor{yellow-dark!17}1.640 \\
DeepSeek-Coder    & \cellcolor{red-dark!19}1.599 & \cellcolor{orange-dark!1}1.849 & \cellcolor{yellow-dark!1}1.868 \\
Qwen2.5-Coder     & \cellcolor{red-dark!0}2.363 & \cellcolor{orange-dark!0}2.486 & \cellcolor{yellow-dark!0}2.338 \\
Qwen3-Coder       & \cellcolor{red-dark!11}1.689 & \cellcolor{orange-dark!31}1.599 & \cellcolor{yellow-dark!16}1.646 \\
\bottomrule
\end{tabular}
\caption{Analysis of interaction efficiency using ATC across the three ambiguity types. The metric measures the average number of dialogue turns required for an agent to successfully recover a missing premise. (Note: Lower values for this metric indicate better efficiency, signifying that the agent secures necessary clarifications with fewer user interactions.)}
\label{tab:14}
\end{table*}

\end{document}